\documentclass[12pt]{article}
\usepackage{graphicx}
\usepackage{amssymb}
\usepackage{amscd}
\begin{document}
\begin{titlepage}

%\begin{center}
%{\hbox to\hsize{
%\hfill \bf hep-ph/??? }}
{\hbox to\hsize{\hfill August 2012 }}

\bigskip \vspace{3\baselineskip}

\begin{center}
{\bf \large 
Standard Model with a distorted Higgs sector and the enhanced Higgs diphoton decay rate}

\bigskip

\bigskip

{\bf Archil Kobakhidze \\ }

\smallskip

{ \small \it
ARC Centre of Excellence for Particle Physics at the Terascale, \\
School of Physics, The University of Sydney, NSW 2006, Australia \\
E-mail: archilk@physics.usyd.edu.au
\\}

\bigskip
 
\bigskip

\bigskip

{\large \bf Abstract}

\end{center}
\noindent 
 We propose a theoretical justification for the anomalous Higgs couplings without extending the particle content of the Standard Model, but rather assuming different realization of the electroweak symmetry and the representation of the Higgs field. Namely, the electroweak  symmetry in our model is non-linearly realised with the Higgs field residing in the singlet representation of the electroweak gauge group.  Within this framework we identify the simplest scenario with CP-violating Higgs-top coupling which explains the enhanced rate of $h\to\gamma\gamma$, while the Higgs production cross section is unaffected.  Remarkably, this can be achieved with a reduced strength of the Higgs-top Yukawa coupling, $\vert y_t\vert<\vert y^{\rm SM}_t\vert$, and, thus, the electroweak vacuum stability problem may also be resolved in our scenario without invoking new physics below the Planck scale. In addition, models with the distorted Higgs sector provide a new framework for potentially successful electroweak baryogenesis.

\end{titlepage}

\paragraph{Introduction.} 
Any theoretically consistent electroweak theory must incorporate:
\begin{itemize}
\item[(i)]  Gauge invariance for massive electroweak gauge bosons;  
\item[(ii)] A mechanism that unitarizes processes involving longitudinal polarization states of 
the massive gauge bosons.
\end{itemize}
The gauge invariance is a necessary ingredient for the consistent description of interacting spin-1 particles (massive and massless). The second requirement is partially related to the first one, since any gauge invariant (and, thus, ghost-free) description of a system of massive vector bosons  with Hermitian Hamiltonian automatically leads to a unitary theory. What we will be assuming here is that there is a mechanism that ensures unitarity of the theory already at the perturbative  level,   without resorting to the full non-perturbative theory. 

The most elegant and simple theory that satisfies the above conditions is the one of spontaneous electroweak symmetry breaking which is implemented within the Standard Model  \cite{Glashow:1961tr, Weinberg:1967tq} through the Higgs mechanism \cite{Higgs:1964ia, Englert:1964et}. In the minimal framework it predicts the existence of an electrically neutral scalar particle, known as the Higgs boson \cite{Weinberg:1967tq, Higgs:1964ia}.  It is rather remarkable therefore, that the recently discovered particle with mass $m_h\approx 125-126$ GeV \cite{:2012gk}  closely reassembles the Higgs boson.   

However, currently available data also show some deviations from the Standard Model predictions. Most notably, $h\to \gamma\gamma$ event rate is by a factor of $1.5 - 2$ larger than its Standard Model value, while $h\to WW, ZZ$ event rates are more consistent with the Standard Model rates. Although the observed enhancement in $h\to \gamma\gamma$ is not yet statistically significant, it may still be hinting at an interesting physics beyond the Standard Model. Therefore, it is important to assess theoretical models that can accommodate these deviations. 

Since $h\to \gamma\gamma$ is a 1-loop radiative process, it is rather sensitive to new charged particles running in the loop. If we require that the Higgs production rate (which is dominated by the radiative gluon fusion process with a virtual top-quark in the loop) is largely unaffected by the new physics, then first natural option is to consider hypotetical charged particles which are colourless. A number of different models along this line has been recently proposed \cite{Dawson:2012di}. The contribution of some coloured states can also be made consistent with the Standard Model Higgs production rate by adjusting corresponding coupling constants, and they also can provide enhancement of the Higgs diphoton rate \cite{Dorsner:2012pp}. We note a potential trouble with such models due to the excessive contribution from new particles to the electroweak precision observables.   

In this paper we propose a theoretical justification for the anomalous Higgs couplings within the framework of distorted Higgs sector. Instead of introducing new particles and symmetries, we assume an alternative realization of the   $SU(2)\times U(1)$ gauge symmetry and a different representation for the Higgs boson. In particular in our approach the electroweak gauge invariance is realized non-linearly with the triplet of nonlinear fields $\pi^ a$ (the would-be Goldstone bosons) parametrizing the coset $SU(2)\times U(1)/U(1)_{\rm EM}$. Obviously, the condition (i) is still satisfied.   The condition (ii) implies that the Higgs boson couples to the electroweak gauge bosons the same way as in the Standard Model. So, the Higgs-gauge boson interactions seems are arrange as if the Higgs boson $\rho$ forms an $SU(2)\times U(1)$-doublet irreducible representation together with the nonlinear Goldstone fields $\pi^a$.  Nevertheless, to specify the representation of $SU(2)\times U(1)$ symmetry group for the Higgs field one must also specify Higgs-fermion interactions and Higgs self-interactions. In fact, with non-linear realization of $SU(2)\times U(1)$ electroweak gauge invariance the Higgs field $\rho$ is no longer obliged to form the electroweak doublet irreducible representation. 
We entertain the possibility that the Higgs field is a singlet of $SU(2)\times U(1)$. As a consequence, a number of new interactions are become possible in the Higgs-fermion sector of the theory. The general (renormalizable) Higgs-Yukawa Lagrangian and the Higgs potential read:  
\begin{eqnarray}
{\cal L}_{HY}=\left[\tilde M^{\rm (u)}_{ij}+Y^{\rm (u)}_{ij}\rho/\sqrt{2}\right]\bar Q^{i}{\cal X}u_{\rm R}^{j}+\left[\tilde M^{\rm (d)}_{ij}+Y^{\rm (d)}_{ij}\rho/\sqrt{2}\right]\bar Q^{i}\tilde{{\cal X}}d_{\rm R}^{j}+\\
\left[\tilde M^{\rm (l)}_{ij}+Y^{\rm (l)}_{ij}\rho/\sqrt{2}\right]\bar L^{i}{\cal X}e_{\rm R}^{j} \nonumber
~,   \label{1}
\end{eqnarray}
\begin{equation}
V=\frac{\mu^2}{2}\rho^2+\frac{\sigma}{3} \rho^3+\frac{\lambda}{4}\rho^4~,
\label{2}
\end{equation}      
where ${\cal X}={\rm e}^{-i\pi^a(x)\tau^{a}+i\pi^3{\bf 1}}(0,1)^{\rm T}$ and $\tilde{{\cal X}}=2i\tau^2{\cal X}^{*}$ [$\tau^a$ ($a=1,2,3$) are the half-Pauli matrices and ${\bf 1}={\rm diag}(1/2,1/2)$]. We assume that the scalar potential has a global minimum for a non-zero vacuum expectation value of the Higgs field $\rho$: 
\begin{equation}
\langle \rho\rangle = v\approx 246~{\rm GeV}~,
\label{3}
\end{equation}
and the shifted field
\begin{equation}
h(x)=\rho(x) - v
\label{4}
\end{equation}
describes the physical excitation associated with the Higgs particle\footnote{The vacuum expectation value in (\ref{3}) is fixed to the standard value since the Higgs interactions with the electroweak gauge bosons are assumed to be the same as in the Standard Model, i.e., $$
\frac{\rho^2}{2}(D_{\mu}{\cal X})^{\dagger}D^{\mu}{\cal X}~,
$$ where $D_{\mu}$ is an $SU(2)\times U(1)$ covariant derivative.}.  The tree-level mass squared of this particle is given by:
\begin{equation}
m_h^2=\left.\frac{\partial^ 2 V}{\partial \rho\partial \rho}\right\vert_{\rho=v}~.
\label{4}
\end{equation} 
According to recent reports \cite{:2012gk},\cite{:2012gu},  the Higgs mass is in the range $m_{h}\approx 125-126$ GeV. 
 
Note that, the cubic interaction term in (\ref{2}) explicitly breaks the discrete $\rho\to -\rho$ symmetry, thus avoiding potentially dangerous cosmological domain wall problem. At the same time, this term may enhance the first order electroweak phase transition needed for a successful electroweak baryogenesis. 

The structure of Higgs-Yukawa Lagrangian (\ref{1}) is quite involved. In the symmetry-breaking phase the physical fermion mass matrices become:
\begin{eqnarray}
M^{\rm (u, d, l)}_{ij}=\tilde M_{ij}^{\rm (u, d, l)}+Y^{\rm (u, d, l)}_{ij}v/\sqrt{2}~.
\label{5}
\end{eqnarray}    
Generally $[\hat M, \hat Y]\neq 0$, and thus new sources of flavour changing processes and CP-violation are present in our approach. It would be interesting to study this generic case in more details and confront its predictions with currently  available data on flavour and CP-violation. 

Remarkably, additional CP-violation appearing beyond the standard Kobayashi-Maskawa CP-violation in our model is also a necessary ingredient for the electroweak baryogenesis. Thus, our model with the distorted Higgs sector provides a new framework where the scenario of the electroweak baryogenesis can potentially be realised. In what follows, we will consider a simplified model with distorted Higgs-top quark sector and show that new CP-violating Higgs-top interactions play crucial role in the enhancement of the Higgs diphoton decay rate, while the Higgs production rate stays the same as in the Standard Model. 

Concluding this section, let us also mention that the electroweak precision observables parametrized through the $S,T,U$ oblique parameters are essentially unaffected in our model. This is because we have not introduced any new particle in our model and the 1-loop oblique parameters depend on particle masses rather than Higgs-Yukawa couplings.

\paragraph{Distorted Higgs-top model.}
Since we are primerely  driven by the potential anomalies in the LHC Higgs data, we consider the simplified model with distorted Higgs-top quark interactions only: 
\begin{eqnarray}
\tilde M^{\rm (d, l)}_{ij}=0,~ \tilde M_{ij}^{\rm (u)}={\rm diag}\left[0,0,\tilde m_t\right]~, \\
Y_{ij}^{\rm (d, l)}=Y_{{\rm SM}~ij}^{\rm (d, l)},~Y_{ij}^{\rm (u)}={\rm diag}\left[ m_u\sqrt{2}/v, m_c\sqrt{2}/v, \tilde y_t\right]~,
\end{eqnarray}    
where we work in the mass eigenstate  basis, and $m_u$, $m_c$ are \emph{up} and \emph{charm} quark masses, respectively. We assume that $\tilde y_t$ coupling is complex, $\tilde y_t=|\tilde y_t|{\rm e}^{i\alpha}$, and hence, top-quark mass is given by: 
\begin{equation}
m_t=\left\vert\tilde m_t +\frac{\tilde y_tv}{\sqrt{2}}\right\vert~.
\end{equation}
 Other Higgs-Yukawa couplings are the same as in the Standard Model, $\hat Y_{\rm SM}$.
 
 Due to the presence of the complex phase $\alpha$, the Higgs-top quark interactions violate CP, as they are comprised of both the CP-even and CP-odd terms:
 \begin{equation}
 {\cal L}_{\rm ht}=\frac{m_t}{v}h\bar t\left[S+i\gamma_5P\right]t~,
 \label{}
 \end{equation}   
where
\begin{eqnarray}
S=\frac{v}{m_t}\sqrt{2}|\tilde y_t|\cos(\alpha)~, \\
A=\frac{v}{m_t}\sqrt{2}|\tilde y_t|\sin(\alpha)~.
\end{eqnarray}

\begin{figure}
\centering
\includegraphics[width=0.8\textwidth]{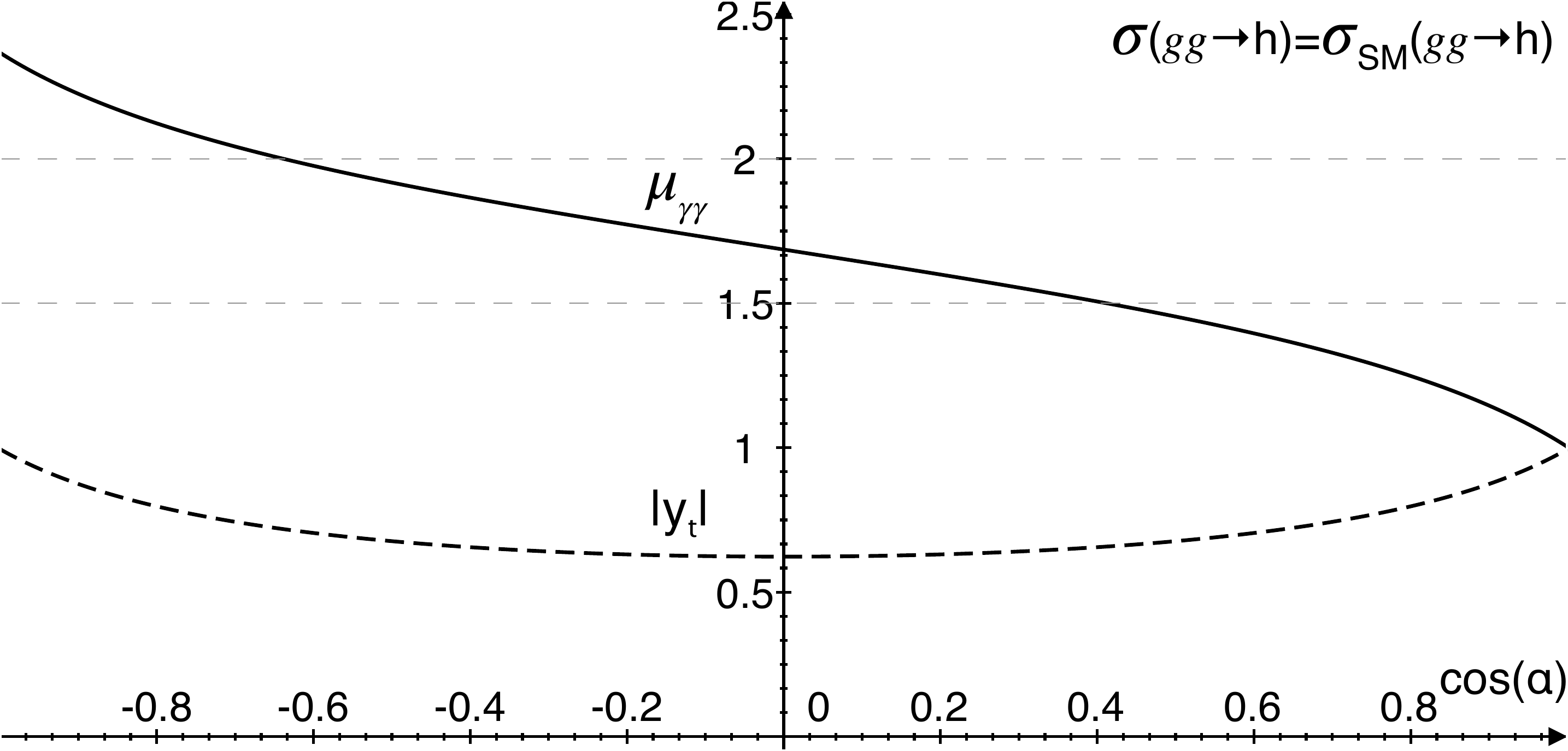}
\vspace{0.5cm}
  \caption{\small The $h\to \gamma\gamma$ signal strength $\mu_{\gamma\gamma}$ (solid curve) and the strength of top-Yukawa coupling $|y_t|$ (dashed curve) as functions of a CP-violating phase. The Higgs production cross section is required to be the same as in the Standard Model. Two horizontal dashed lines show the observational range for $\mu_{\gamma\gamma}$. }
  \label{fig}
\end{figure}    

Let us consider now $h\to \gamma\gamma$ decay process. Contributions of CP-even and CP-odd interactions to this decay process do not interfere. This has twofold effect in enhancing $h\to \gamma\gamma$ decay rate: (i) destructive interference between CP-even $h\bar tt$ interactions and $hW^{+}W^{-}$ interactions is reduced and (ii) CP-odd contribution always enhances the decay rate since it does not interfere with $hW^{+}W^{-}$ contribution. 

Remarkably, due to the extra CP-violation the model allows enhancement of the Higgs diphoton rate, while keeping the Higgs production rate unaffected.  The Higgs production rate is dominated by the gluon fusion process, and in the narrow width approximation we can write,
\begin{equation}
\frac{\sigma(gg\to h)}{\sigma_{\rm SM}(gg\to h)}\approx \frac{\Gamma (h\to gg)}{\Gamma_{SM}(h\to gg)}=1~,
\end{equation} 
as a condition that the Higgs production rate is the same as in the Standard Model. The above equation implies for the strength of the top-Yukawa coupling, $|y_t|=\frac{\sqrt{2}v}{m_t}|\tilde y_t|\approx 2|\tilde y_t|$:
\begin{equation}
|y_t|=\frac{2}{\sqrt{10.31-6.27\cos^2(\alpha)}}~.
\end{equation}
 Using this result, we can find for the $h\to \gamma\gamma$ signal strength $\mu_{\gamma\gamma}= \sigma(gg\to h)\times Br(h\to \gamma\gamma)/\sigma_{\rm SM}(gg\to h)\times Br_{\rm SM}(h\to \gamma\gamma)\approx
\Gamma(h\to \gamma\gamma)/\Gamma_{\rm SM}(h\to \gamma\gamma)$:
\begin{equation}
\mu_{\gamma\gamma}=\left(1.27-\frac{0.54\cos(\alpha)}{\sqrt{10.31-6.27\cos^2(\alpha)}}\right)^2+
\frac{0.754\sin^2(\alpha)}{10.31-6.27\cos^2(\alpha)}~.
\end{equation}     
The last two equations are plotted in Fig.\ref{fig}. These plots show that, depending on the CP-violating phase, our theory interpolates between the usual Standard Model ($\cos(\alpha)=1$) and the Standard Model with opposite sign top-Yukawa coupling ($\cos(\alpha)=-1$), $y_t=-y_t^{\rm SM}$. In the later case $\mu_{\gamma\gamma}\approx 2.4$, and it is slightly  above the current measured values. This enhancement is due to the fact that the destructive interference between $h\bar tt$ and $hW^{+}W^{-}$ contributions in the Standard Model turns into a constructive interference with the opposite sign top-Yukawa coupling \cite{Espinosa:2012ir}. Between these two end-points we have a large range of non-trivial CP-violating phases $\alpha$ where the signal strength is in the observational range, $\mu_{\gamma\gamma}=1.5-2$:
\begin{equation}
-0.64<\cos(\alpha)<0.41~.
\label{asd}
\end{equation}
It would be interesting to study whether the above CP-violating phase is enough to produce also the observed matter-anti-matter asymmetry through the electroweak baryogenesis.

Another remarkable aspect of the model is that the top-Yukawa coupling corresponding to the range in Eq. (16) is smaller  than the Standard Model value, 
\begin{equation}
0.62<|y_t/y_t^{SM}|<0.72~.
\end{equation}
Note that, the lower bound in (17) corresponds to a purely CP-odd coupling ($\cos(\alpha)=0$). Eq. (17) implies  that bounds on the Higgs mass from the electroweak vacuum stability can be significantly relaxed in our model, allowing to have a consistent theory with $m_h=125-126$ GeV up to the Planck scale \cite{new}.

\paragraph{Conclusion and outlook.} We have proposed a new class of models beyond the Standard Model without introducing new particles and symmetries. Instead, we assume that the electroweak gauge symmetry is realised non-linearly and the Higgs boson resides in the singlet representation of the Standard Model gauge group. We have identified a simple model of this class which features a new CP-violating Higgs-top quark coupling and is otherwise indistinguishable from the Standard Model. For a large range of the new CP-violating phase the model successfully reproduces the basic features of the current LHC data: it predicts enhanced Higgs diphoton decay rate, while the Higgs production rate is the same as in the Standard Model. As a bonus features, the proposed class of models provide a new framework for the successful electroweak baryogenesis, and resolves the vacuum stability problem associated with the $125-126$ GeV Higgs boson. 

If the anomalies in current Higgs data will be confirmed in future experiments, further check of the propose model is possible by identifying CP-violating nature of the top-Yukawa coupling through measurements of an angular distribution of the final state photons in $h\to\gamma\gamma$ decays. In addition, future measurements of the Higgs production process in association with top-anti-top pair may reveal the reduced strength of the top-Yukawa coupling (17) predicted in our model.

\paragraph{Acknowledgment.} This work was partially supported by the Australian Research Council.

\end{document}